\documentclass[pra,aps,floatfix,showpacs,tightenlines,twocolumn,
amsmath,amssymb,nofootinbib]{revtex4-1}

\usepackage{epsfig}
\usepackage{subfigure}
\usepackage{amsmath, amssymb, amsthm, amsfonts}
\usepackage{bm}
\usepackage{braket}
\usepackage[english]{babel}

\renewcommand{\vec}[1]{\mathbf{#1}}
\newcommand{\mat}[1]{\mathbf{#1}}
\newcommand{\op}[1]{\hat{#1}}
\newcommand{\ii}{\operatorname{i}}

\newcommand{\defemph}[1]{\emph{#1}}
\newcommand{\transpose}[1]{#1 ^{\operatorname T}}

\newcommand{\half}{\frac{1}{2}}
\newcommand{\dd}{\mathrm{d}}

\begin{document}
\title{Optimal Gaussian Entanglement Swapping}
\author{Jason Hoelscher-Obermaier}
\author{Peter van Loock}\email{peter.vanloock@mpl.mpg.de}
\affiliation{$^1$Optical Quantum Information Theory Group, Max Planck Institute for the Science of Light, G\"unther-Scharowsky-Str. 1/Bau 26, 91058 Erlangen, Germany}
\affiliation{Institute of Theoretical Physics I, Universit\"at Erlangen-N\"urnberg, Staudtstr. 7/B2, 91058 Erlangen, Germany}

\begin{abstract}
We consider entanglement swapping with general mixed two-mode Gaussian states and calculate the optimal gains for a broad class of such states including those states most relevant in communication scenarios. We show that for this class of states, entanglement swapping adds no additional mixedness, that is the ensemble average output state has the same purity as the input states. This implies that, by using intermediate entanglement swapping steps, it is, in principle, possible to distribute entangled two-mode Gaussian states of higher purity as compared to direct transmission. We then apply the general results on optimal Gaussian swapping to the problem of quantum communication over a lossy fiber and demonstrate that, contrary to negative conclusions in the literature, swapping-based schemes in fact often perform better than direct transmission for high input squeezing. However, an effective transmission analysis reveals that the hope for improved performance based on optimal Gaussian entanglement swapping is spurious since the swapping does not lead to an enhancement of the effective transmission. This implies that the same or better results can always be obtained using direct transmission in combination with, in general, less squeezing.
\end{abstract}

\maketitle

\section{Introduction}
Entanglement swapping is a standard tool in quantum information processing \cite{zukowski93,briegel98}. It is particularly relevant in the context of quantum communication where, due to losses in the channel, one typically deals with mixed states. In the case of qubits, the states which naturally arise in quantum communication contexts include mixed Bell-diagonal and Werner states. Entanglement swapping with these states is known to further increase the mixedness of the initial states, e.g.~for a Bell-diagonal state of the form $F \ket{\Phi^+}\bra{\Phi^+} + (1-F)\ket{\Phi^-}\bra{\Phi^-}$, the output state is of the same form, but with decreased fidelity $F'= F^2 + (1-F)^2$. Hence, in this case, the purity decreases exponentially \cite{briegel98} (but see \cite{sen05,modlawska08}). 
Surprisingly, qubit entanglement swapping has nevertheless been found to be of use for quantum key distribution since intermediate swappings can mitigate the detrimental effects of detector dark counts on the security of quantum key distribution schemes \cite{collins05}. 

Given the known practical benefits of Gaussian quantum information processing \cite{eisert03,adesso07} such as the possibility of deterministic entanglement generation and swapping \cite{braunstein05}, it is quite surprising that entanglement swapping with mixed Gaussian states has not yet been treated comprehensively, for example, with an eye on quantum key distribution or other quantum communication tasks. In particular, quantum communications over large distances using quantum repeaters \cite{briegel98} would have to rely on entanglement swapping as one of the main elementary steps.

In Ref.~\cite{pirandola06}, entanglement swapping with general mixed Gaussian states has indeed been considered and the output states for the single-shot case have been derived. For applications in experiments, however, it is important to analyze the ensemble average case in which, contrary to the single-shot case, the output state is taken to be a mixture of conditional output states with different displacements depending on the Bell measurement result. In the ensemble average case, the conditional displacements have to be undone in order to optimize the output state. In particular, the gains for the experimentally applied displacements have to be chosen in the right way depending on the input state. 

Refs.~\cite{liu02,ban04,li05}, on the other hand, analyzed the gain-dependent ensemble output states for swapping with mixed Gaussian states but failed to identify the optimal swapping scheme and therefore also failed to provide a basis for any general negative conclusion.\footnote{So far, optimal entanglement swapping has only been discussed for the special case of pure two-mode squeezed states \cite{loock02}, which is insufficient for the application to quantum communication.} More specifically, references \cite{liu02,ban04,li05} use only one-sided displacements which is in general not optimal. To achieve optimal results, two-sided displacements have to be used. In fact, we demonstrate that, for high values of the input squeezing, transmission with an intermediate swapping step often yields better output states (in terms of the EPR correlations \cite{einstein35}, as well as the entanglement of formation) than direct transmission --- contrary to the negative conclusion in \cite{liu02} in which it is claimed that swapping always performs worse than direct transmission. This is intuitively reasonable: first, in analogy with the pure-state case, we may expect swapping to perform better and better for higher values of the input squeezing. Second, the sensitivity to losses increases with increasing squeezing. Therefore, the fact that in the swapping-based scheme each of the states incurs only half as much loss as in direct transmission can be expected to yield an advantage for the swapping-based scheme.

Nevertheless, the apparent gain in the performance due to entanglement swapping in the high-squeezing regime turns out to be spurious. In our examples, even better results are obtained if direct transmission is used together with less input squeezing. This statement can be made rigorous and general using an effective loss analysis which reveals that the intermediate swapping step does never increase the transmittivity of the quantum channel. This result implies that, in general, direct transmission performs at least as good as swapping-based schemes and any supposed advantage of the swapping-based schemes disappears if the optimal input squeezing is used.

The structure of the paper is as follows. In section~\ref{sec:prelims}, we introduce some basic concepts of Gaussian quantum information theory. In section~\ref{sec:swapping}, we discuss entanglement swapping with mixed two-mode Gaussian states and calculate the conditional output states, as well as the ensemble average output states. In section~\ref{sec:output_analysis}, we analyse the ensemble average output state and derive the optimal gains for a large class of input states. We show that exactly if the optimal gains are used, the ensemble average output state is given by the conditional output state. We argue that this choice of the gains yields the optimal output entanglement, optimal output EPR correlations, as well as optimal output purity. In fact, the purity of the optimal ensemble-average output state is equal to the purity of the input states, hence the optimal swapping preserves the purity of the input states. In section~\ref{sec:qucomm}, we apply the results of section~\ref{sec:output_analysis} to the problem of quantum communication over a lossy fiber. We demonstrate that there are values of the input squeezing, for which a swapping-based communication scheme is able to distribute more entanglement of formation and better EPR correlations than direct transmission. In section~\ref{sec:effloss}, we use an effective loss analysis to show that swapping does not lead to an increased transmittivity. Therefore, the same or better output states can be obtained by direct transmission if different input squeezing is used. Finally, section \ref{sec:discussion} provides a summary of our results.

\section{Preliminaries}
\label{sec:prelims}
For a single bosonic mode with annihilation and creation operators $\op a, \op a ^\dagger$, the \defemph{quadrature operators} $\op q , \op p $ are defined by
\begin{equation}
\label{eq:quadratures}
 \op q = \frac{1}{\sqrt 2} \left( \op a + \op a ^\dagger \right), \quad \op p = \frac{1}{\sqrt 2 \ii} \left( \op a - \op a ^\dagger \right) .
\end{equation}
The $q$ and $p$ without operator hats denote the corresponding \defemph{quadrature variables} which are the arguments of the Wigner function corresponding to a given quantum state. When dealing with quantum states of multiple modes it is convenient to define the vector of quadrature variables for an $n$-mode quantum state as
\begin{equation}
\vec{R}_{i_1,\ldots ,i_n} = (q_{i_1},p_{i_1},\ldots, q_{i_n},p_{i_n}) , \nonumber
\end{equation}
and, similarly, for the vector of quadrature operators defined as $\op{\vec{R}}_{i_1,\ldots ,i_n}$. $\braket{\op X} $ denotes the expectation value (first moment) of the operator $\op X$ in the quantum state in question. Hence, $\braket{ \op{\vec{R}}_{i_1,\ldots ,i_n} }$ denotes the vector of first moments.

The \defemph{covariance matrix} (CM) of a two-mode state $\rho_{AB}$ on modes $A$ and $B$ is a real symmetric $4 \times 4$ matrix ${\mat\sigma_{AB}}$ with entries
\begin{equation}
 {\left( \mat\sigma_{AB} \right)}_{ij} =
 \half \braket{ \op{\vec{R}}_{i} \op{\vec{R}}_{j} + \op{\vec{R}}_{j} \op{\vec{R}}_{i}}
 - \braket{\op{\vec{R}}_{i}}\braket{\op{\vec{R}}_{j}} ,
\end{equation}
where $\op{\vec{R}}$ is short for $\op{\vec{R}}_{AB}$.

Note that not every real symmetric $4 \times 4$ matrix is the CM of a quantum state since the variances have to fulfill the Robertson-Schr\"odinger inequality \cite{simon00}.

\defemph{Gaussian states} are states whose Wigner function is a Gaussian distribution of the quadrature variables. They are completely characterized by the first and second moments of the quadrature operators, i.e.,~by $\braket{ \op{\vec{R}}_{AB} }$ and ${\mat\sigma_{AB}}$.

By local Gaussian operations the first moments of any two-mode Gaussian state can be set to zero and the CM can be brought into the following standard form~\cite{simon00,duan00},
\begin{equation}
\mat{\sigma} = \begin{pmatrix}
a &  & c_+ &  \\
 & a &  & c_- \\
c_+ &  & b &  \\
 & c_- &  & b
\end{pmatrix}
.
\label{eq:simonCM}
\end{equation}

\section{Entanglement Swapping}
\label{sec:swapping}
Let us now consider entanglement swapping with two identical Gaussian states with CM \eqref{eq:simonCM} and vanishing first moments on modes 1 and 2 and modes 3 and 4, i.e.,
\begin{equation}
{\mat\sigma_{12}}={\mat\sigma_{34}} = \mat{\sigma} , \quad \braket{\op{\vec{R}}_{1,2}} = \braket{\op{\vec{R}}_{3,4}} = \vec 0.
\end{equation} In this case, the Wigner function of the 4-mode system before the swapping is given by
\begin{eqnarray}
W_\text{in} (\vec{R}_{1,2,3,4}) &=& W_\text{in}(\vec{R}_{1,2}) W_\text{in}(\vec{R}_{3,4}) ,    \label{eq:wignerIn}  \\
W_\text{in}(\vec{R}_{i,j})  &=& \mathcal N^{-1} \exp\left( - 1/2 \; \vec{R}_{i,j} \mat{\sigma}^{-1} \transpose{\vec{R}_{i,j}} \right) \nonumber ,
\end{eqnarray}
where $\mathcal N =  \int\ldots\int  \exp\left( - 1/2 \; \vec{R}_{i,j} \mat{\sigma}^{-1} \transpose{\vec{R}_{i,j}} \right) \dd^4 \vec{R}_{i,j}$ is a normalization factor.

The swapping is performed via a Bell measurement on modes 2 and 3 \cite{braunstein98}. First, these modes are combined at a 50:50 beam splitter yielding outgoing modes $u$ and $v$ which are described by the annihilation operators
\begin{equation}
 \hat a_u = \frac{1}{\sqrt 2} \left( \hat a_2 - \hat a_3 \right), \qquad
 \hat a_v = \frac{1}{\sqrt 2} \left( \hat a_2 + \hat a_3 \right) .
\end{equation}

Hence the Wigner function $W_\text{BS} (\vec{R}_{1,u,v,4)}$ 
of the state after the beam splitter as a function of the new quadratures $q_u,p_u,q_v,p_v$ is obtained by substituting the old variables $q_2,p_2,q_3,p_3$ in the Wigner function \eqref{eq:wignerIn} as follows,
\begin{eqnarray}
q_2  \rightarrow \frac{q_u+q_v }{\sqrt{2}}, \quad
p_2  \rightarrow \frac{p_u+p_v }{\sqrt{2}},  \nonumber \\
q_3  \rightarrow \frac{q_v -q_u }{\sqrt{2}}, \quad
p_3  \rightarrow \frac{p_v -p_u }{\sqrt{2}}.
\end{eqnarray}

Then, the new quadratures $q_u$ and $p_v$ are measured using homodyne detection. If we denote the measurement results  by $q_u'$ and $p_v'$, the Wigner function of modes 1 and 4 after the measurement is obtained by integrating $W_\text{BS}$ over the unmeasured quadratures $q_v $ and $p_u$ and setting $q_u=q_u'$ and $p_v=p_v'$:
\begin{eqnarray}
\label{eq:wignerConditional}
W_\text{cond} (\vec{R}_{14}) &&= \\
&&\left. \iint \dd q_v \dd p_u W_\text{BS}  (\vec{R}_{1,u,v,4}) \right\arrowvert_{q_u=q_u',p_v=p_v'} . \nonumber
\end{eqnarray}

Note that the above Wigner function $W_\text{cond}$ for the conditional state is unnormalized. Its norm is the probability (density) $\mathrm{pr} \left( q_u' , p_v'\right)$ that the Bell measurement yields the results $q_u=q_u'$ and $p_v=p_v'$. Depending on the CM of the input states, the conditional state after the Bell measurement is an entangled state on modes 1 and 4 modulo a displacement which is determined by the Bell measurement result, as will be shown below. The CM of the conditional state can be calculated by integration from its Wigner function \eqref{eq:wignerConditional}. It is independent of the Bell measurement results $q_u'$ and $p_v'$ and given by
\begin{equation}
 \label{eq:conditionalCM}
\mat{\sigma}_\text{cond}=
\left(
\begin{array}{cccc}
 a-\frac{c_+^2}{a+b} &  & \frac{c_+^2}{a+b} &  \\
  & a-\frac{c_-^2}{a+b} &  & -\frac{c_-^2}{a+b} \\
 \frac{c_+^2}{a+b} &  & b-\frac{c_+^2}{a+b} &  \\
  & -\frac{c_-^2}{a+b} &  & b-\frac{c_-^2}{a+b}
\end{array}
\right).
\end{equation}

Hence, all conditional output states are states with the same CM \eqref{eq:conditionalCM}, but with different first moments depending on $q_u'$ and $p_v'$.\footnote{The first moments are proportional to the coefficients of the linear term in the exponent of the Wigner function of the conditional state given by Eq.~\eqref{eq:wignerConditional}. These coefficients depend on the Bell measurement results $q_u'$ and $p_v'$ and read
\begin{equation}
\sqrt{2} (a + b) (a b - c_-^2) (a b - c_+^2)
\begin{pmatrix}
 q_u'  b  c_+   \left(a b-c_-^2\right) \\
 p_v'  b  c_-   \left(a b-c_+^2\right) \\
 q_u'  a  c_+   \left(c_-^2-a b\right) \\
 p_v'  a  c_-   \left(a b-c_+^2\right)
\end{pmatrix}.
\end{equation}}

Due to the randomness of the Bell measurement results, one typically deals with an entangled output state subject to continuously fluctuating displacements~\cite{loock02}. These displacements have to be undone by displacing modes $1$ and $4$ according to the Bell measurement results $ q_u'$ and $ p_v'$.  In order to obtain optimal output states, the displacements have to be weighted by \defemph{gain factors} which will be denoted by $g_1$ and $g_4$ for the displacement of modes 1 and 4, respectively.
Modes $1$ and $4$ are displaced by $\sqrt{2} g_1(-q_u' + \ii p_v')$ and $\sqrt{2} g_4(+q_u' + \ii p_v')$, respectively~\cite{loock02}, such that the output quadratures read
\begin{align}
\hat q_{1\text{out}} &=\hat q_1 - g_1 \sqrt 2  q_u' ,\quad \hat p_{1\text{out}}=\hat p_1 + g_1 \sqrt 2  p_v' , \nonumber\\
 \hat q_{4\text{out}}&=\hat q_4 + g_4 \sqrt 2  q_u' ,\quad \hat p_{4\text{out}}=\hat p_4 + g_4 \sqrt 2  p_v' .
\label{eq:CV_quantum_relay_output_modes_single-shot}
\end{align}

The Wigner function $W_\text{displ} (\vec{R}_{1\text{out},4\text{out}})$ for the conditional output state after the displacements can be obtained by substituting the new variables $q_{1\text{out}},\ldots ,p_{4\text{out}}$ for the old variables $q_1,\ldots,p_4$ according to Eq.~\eqref{eq:CV_quantum_relay_output_modes_single-shot} in the Wigner function $W_\text{cond}$ of the conditional state in Eq.~\eqref{eq:wignerConditional}.
The displacement does not change the CM of the conditional state which is therefore still given by Eq.~\eqref{eq:conditionalCM}. For general gains $g_1$ and $g_4$ however, the displaced conditional states still differ in their first moments. The first moments are proportional to the coefficients of the linear terms of the exponent of the Wigner function $W_\text{displ}$, given by
\begin{equation}
\label{eq:linTermsGeneral}
\sqrt{2}
\begin{pmatrix}
	-\frac{q_u \left(g_1 b^2+(a   g_1  -c_+) b+c_+^2   (g_4-g_1)\right) }{(a+b) \left(a   b-c_+^2\right)} \\
	\frac{p_v \left(g_1   b^2+(c_-  +a g_1) b+c_-^2   (g_4-g_1)\right) }{(a+b) \left(a   b-c_-^2\right)} \\
	\frac{q_u \left(g_4   a^2+(b g_4- c_+) a+c_+^2   (g_1-g_4)\right) }{(a+b) \left(a   b-c_+^2\right)} \\
	\frac{p_v \left(g_4   a^2+(c_-  +b g_4) a+c_-^2   (g_1-g_4)\right) }{(a+b) \left(a   b-c_-^2\right)}
\end{pmatrix}.
\end{equation}

In order to obtain the Wigner function $W_\text{ens}$ of the output state in the ensemble-average case, one has to average the Wigner function $W_\text{displ}$ of the displaced conditional states over all possible measurement results $ q_u'$ and $ p_v'$ weighted by the probability density $\mathrm{pr} \left( q_u' , p_v'\right)$ for the respective result. This yields a two-mode Gaussian state with vanishing first moments and CM \newline
\begin{widetext}
\begin{equation}
\label{eq:ensembleCMgeneral}
\mat{\sigma}_\text{ens}=
\left(
\begin{array}{llll}
a +  (a+b) g_1^2-2 c_+ g_1   &      & c_+   (g_1+g_4) - g_1 g_4 (a+b)  &  \\
  & a + (a+b) g_1^2+2 c_- g_1 &  & c_-   (g_1+g_4) + g_1 g_4 (a+b) \\
c_+ (g_1+g_4) - g_1 g_4(a+b) &  & b + (a+b) g_4^2-2   c_+ g_4 &  \\
  & c_- (g_1+g_4) + g_1 g_4 (a+b) &  & b + (a+b) g_4^2+2   c_- g_4
\end{array}
\right).
\end{equation}
\end{widetext}

\section{Analysis of the Output State}
\label{sec:output_analysis}
We saw that, for any two input states of the form \eqref{eq:simonCM}, all conditional output states have the same CM \eqref{eq:conditionalCM}. Note that, for Gaussian states, the most interesting properties such as the entanglement, the quality of the EPR correlations, and the purity, only depend on the CM. Hence, all conditional output states have the same entanglement $E_\text{cond}$, EPR correlations $\Delta\text{EPR}_\text{cond}$, and purity $\mu_\text{cond}$. Intuitively, the ensemble average over these conditional states with the same CM~\eqref{eq:conditionalCM} but different first moments~\eqref{eq:linTermsGeneral} leads to a ``blurring out'' of the variances and covariances, yielding an output state $\rho_\text{ens}$ of less quality.

More formally, it follows for any entanglement monotone~$E$ which is convex, such as the entanglement of formation, that the entanglement~$E\left(\rho_\text{ens}\right)$ of the ensemble average state is bounded from above by the entanglement~$E_\text{cond}$ of the conditional states:
\begin{eqnarray}
E \left( \rho_\text{ens} \right) &=&
E \left( \iint \dd q_u' \dd p_v' \mathrm{pr} \left( q_u' , p_v'\right) \rho_{\text{cond},q_u' p_v'} \right)\nonumber \\
&\leq&  \iint \dd q_u' \dd p_v' \mathrm{pr} \left( q_u' , p_v'\right) E \left(  \rho_{\text{cond},q_u' p_v'} \right) \nonumber \\
&=& \iint \dd q_u' \dd p_v' \mathrm{pr} \left( q_u' , p_v'\right) E_\text{cond} = E_\text{cond}. 
\end{eqnarray}
The conditional states $\rho_{\text{cond},q_u' p_v'}$ are here assumed to be normalized.

A very similar argument shows that the purity of the ensemble average state $\mu \left( \rho_\text{ens} \right)$ is bounded from above by the purity of the conditional states $\mu_\text{cond}$. For simplicity, let us denote the Bell measurement results collectively as $\left( q_u', p_v' \right) = \vec{m}$ and the conditional output state $\rho_{\text{cond},\vec{m}}$ by $\rho_{\vec{m}}$. Then, the purity of the ensemble average state is given by
\begin{eqnarray}
&&\mu\left( \rho_\text{ens} \right) = \mathrm{tr} \left(\rho_\text{ens}^2 \right) \\
&=& \mathrm{tr} \left( \int\ldots\int \dd^2 \vec{m}_1 \dd^2 \vec{m}_2
\mathrm{pr} \left(\vec{m}_1 \right) \mathrm{pr} \left(\vec{m}_2 \right)
\rho_{\vec{m}_1}\rho_{\vec{m}_2}
 \right) \nonumber\\ 
&=&  \int\ldots\int \dd^2 \vec{m}_1 \dd^2 \vec{m}_2
 \mathrm{pr} \left(\vec{m}_1 \right) \mathrm{pr} \left(\vec{m}_2 \right)
\mathrm{tr} \left(
 \rho_{\vec{m}_1}\rho_{\vec{m}_2}
  \right) \nonumber\\
&\leq&  \int\ldots\int \dd^2 \vec{m}_1 \dd^2 \vec{m}_2
   \mathrm{pr} \left(\vec{m}_1 \right) \mathrm{pr} \left(\vec{m}_2 \right)
  \mathrm{tr} \left(
   \rho_{\vec{m}_1}^2
    \right) \nonumber\\ 
&=&  \int\ldots\int \dd^2 \vec{m}_1 \dd^2 \vec{m}_2
   \mathrm{pr} \left(\vec{m}_1 \right) \mathrm{pr} \left(\vec{m}_2 \right)
 \mu_\text{cond} \nonumber\\
 &=& \mu_\text{cond} . \nonumber
\end{eqnarray}
In the step from line 3 to line 4, we used
$\mathrm{tr} \left(
   					\rho_{\vec{m}_1}\rho_{\vec{m}_2}
		    \right)
\leq     \mathrm{tr} \left(
    						\rho_{\vec{m}_1}^2
      		\right)$.
This follows from the Cauchy-Schwarz inequality applied to the Hilbert-Schmidt norm which yields
$
\left(
\mathrm{tr} \left(
   					\rho_{\vec{m}_1}\rho_{\vec{m}_2}
		    \right)  \right)^2
\leq     \mathrm{tr} \left(
    						\rho_{\vec{m}_1}^2
      		\right)\mathrm{tr} \left(
      		    						\rho_{\vec{m}_2}^2
      		      		\right)
$, together with the fact that
      		      		$\mathrm{tr} \left(
      		      		    						\rho_{\vec{m}_1}^2
      		      		      		\right)= \mathrm{tr} \left(
      		      		      		    						\rho_{\vec{m}_2}^2
      		      		      		      		\right) = \mu_\text{cond}$.

Finally, it is easy to see that the variance of the EPR quadratures is bounded from below by the variance of the EPR quadratures of the conditional output states $\Delta\text{EPR}_\text{cond}$. %
For a state $\rho$ with vanishing first moments, the EPR correlations are given by
\begin{eqnarray}
\Delta\text{EPR} ( \rho ) &=&  \mathrm{tr} \left( \hat E \rho \right), \\
 \hat E  &:=& 1/2 \left( \left( \hat q_1 - \hat q_4 \right)^2 + \left( \hat p_1 + \hat p_4 \right)^2 \right). \nonumber
\end{eqnarray}
Note first that the quantity $\Delta\text{EPR}$ as defined above is not invariant under local (Gaussian) unitary operations. To obtain an interesting figure of merit which can be a basis for a comparison of different quantum communication schemes, one should optimize the EPR correlations over all possible local Gaussian operations $U_1,U_4$:\footnote{The optimal EPR correlations are interesting since they are directly related to the fidelity that can be achieved in quantum teleportation \cite{braunstein98,furusawa98}.}
\begin{equation}
\label{eq:EPRopt}
\Delta\text{EPR}_\text{opt} ( \rho ) := \min\limits_{U = U_1 \otimes U_4} \left( \mathrm{tr} \left( \hat E U \rho U^\dagger \right) \right).
\end{equation}
Using the linearity of the trace, we obtain
\begin{eqnarray}
&&\Delta\text{EPR}_\text{opt} \left( \rho_\text{ens} \right)=
\min\limits_{U = U_1 \otimes U_4}
\left( \mathrm{tr} \left( \hat E U  \rho_\text{ens} U^\dagger \right)  \right)\\
&&= \min 
		\left( \iint \dd q_u' \dd p_v' \mathrm{pr} \left( q_u' , p_v'\right) \mathrm{tr} \left( \hat E U  \rho_{\text{cond},q_u' p_v'} U^\dagger \right)  \right)
 \nonumber \\
&&\geq \iint \dd q_u' \dd p_v' \mathrm{pr} \left( q_u' , p_v'\right)  \min 
 \left( \mathrm{tr} \left( \hat E U  \rho_{\text{cond},q_u' p_v'} U^\dagger \right)  \right) \nonumber\\
&&= \iint \dd q_u' \dd p_v' \mathrm{pr} \left( q_u' , p_v'\right)  \Delta\text{EPR}_\text{opt} \left( \rho_{\text{cond},q_u' p_v'} \right)  \nonumber \\
&&= \Delta\text{EPR}_\text{opt} \left( \rho_{\text{cond}} \right),
\end{eqnarray}
where the last equality follows since the second moments of $\rho_{\text{cond},q_u' p_v'}$ are independent of $q_u'$ and $ p_v'$ and $\Delta\text{EPR}_\text{opt} \left( \rho_{\text{cond},q_u' p_v'} \right)$ only depends on the second moments. Hence, no matter which of the considered quantities we are interested in, the conditional output state is the best we can achieve.

However, only if the displacements of the conditional states given by Eq.~\eqref{eq:linTermsGeneral} vanish (due to the right choice of the gains $g_1,g_4$), is the CM \eqref{eq:ensembleCMgeneral} of the ensemble average state equal to that of the conditional state in Eq.~\eqref{eq:conditionalCM}.
This can be achieved if and only if $c_+ = -c_- =:c$.%
\footnote{This statement relies on the assumption of phase-independent gains as in Eq.~\eqref{eq:CV_quantum_relay_output_modes_single-shot}. If the gains are allowed to depend not only on the mode, but also on the quadrature such that the quadratures after the displacement are given by
\begin{eqnarray}
 q_{1,\text{out}}= q_1 - \sqrt{2} g_{1,q} q_u, \quad  p_{1,\text{out}} = p_1 + \sqrt{2} g_{1,p}  p_v, \nonumber\\
q_{4,\text{out}} = q_4 + \sqrt{2} g_{4,q}  q_u, \quad   p_{4,\text{out}} = p_4 + \sqrt{2} g_{4,p} p_v,
\end{eqnarray}
one can indeed achieve an ensemble-average output state which is equal to the single-shot output state also in the more general case in which $c_+  \neq  -c_-$. In this case, the optimal quadrature-dependent gains are given by
\begin{equation}
g_{1,q} = g_{4,q}= c_+/(a + b), \quad g_{1,p} = g_{4,p}= -(c_-/(a + b)).
\end{equation}
For simplicity, we will stick to the simpler case of phase-independent gains in the main text.}
 In this case, the displacement vanishes for all measurement results if we set
\begin{equation}
 g_1 =g_4= \frac{c}{a+b} .
 \label{eq:opt_sym_gains}
\end{equation}
Let us summarize at this point. If the input states are described by a CM of the form
\begin{equation}
\label{eq:simpleCM}
\begin{pmatrix}
a & & c & \\
& a & & -c \\
c & & b & \\
& -c & & b
\end{pmatrix},
\end{equation}
then the gains can be chosen according to Eq.~\eqref{eq:opt_sym_gains} such that the output state in the ensemble-average case is optimal, namely equal to the conditional output state for $q_u' = p_v' = 0$. The resulting two-mode Gaussian state has vanishing first moments and a CM
\begin{equation}
\label{eq:optimalOutCM}
\mat{\sigma}_\text{opt}=
\left(
\begin{array}{cccc}
 a-\frac{c^2}{a+b} &  & \frac{c^2}{a+b} &  \\
  & a-\frac{c^2}{a+b} &  & -\frac{c^2}{a+b} \\
 \frac{c^2}{a+b} &  & b-\frac{c^2}{a+b} &  \\
  & -\frac{c^2}{a+b} &  & b-\frac{c^2}{a+b}
\end{array}
\right).
\end{equation}

Note also that, for any input state of the form \eqref{eq:simonCM}, the purity of the resulting conditional state \eqref{eq:conditionalCM} is given by $1/\left(4 \sqrt{(a b - c_-^2) (a b - c_+^2)}\right)$ which is equal to the purity of the input state. %
This implies the following remarkable fact: \emph{for mixed two-mode Gaussian states, the purity can be completely preserved during the entanglement swapping.} In particular, for states of the form \eqref{eq:simpleCM}, one can maintain the input purity even for the ensemble average output states. This result extends earlier work which showed that for entanglement swapping with pure finitely squeezed two-mode squeezed states, the gains can be chosen such that the (ensemble average) output state is pure again, i.e.,~such that the purity is preserved in the entanglement swapping step \cite{loock02}.

Note that one of the most widely used entanglement monotones, the logarithmic negativity, fails to be convex \cite{eisert01,vidal02,plenio05}. Therefore one cannot automatically assume that the logarithmic negativity of the ensemble average output state is bounded from above by that of the conditional output state. Instead, one has to explicitly optimize the logarithmic negativity of the ensemble average output state over all possible gains. We find that, for states of the form \eqref{eq:simpleCM}, the logarithmic negativity as a function of the gains $g_1$ and $g_4$ is critical along the following path in the $g_1$-$g_4$ parameter space,
\begin{eqnarray}
\label{eq:optGainPath}
&&\vec{\gamma}(t):= (g_{1,\text{critical}} (t), g_{4,\text{critical}} (t) ) = \\
&& \left( \frac{c}{a+b}+\left(a^2-b^2 + \sqrt{4 c^4+\left(a^2-b^2\right)^2}\right) t,\frac{c}{a+b}-2 c^2 t\right). \nonumber
\end{eqnarray}
Further analysis reveals that, for all parameters $a,b,c$ for which the input states are physical and entangled, the second derivative of the logarithmic negativity perpendicular to the critical path is negative. Since the points on the path are the only critical points, this establishes that 
the gains on the path optimize the output logarithmic negativity. In particular, the optimal gains include the gains $g_1= g_4 =c/(a+b)$. Hence, for states of the form \eqref{eq:simpleCM}, the gains \eqref{eq:opt_sym_gains} which yield an ensemble average output state \eqref{eq:ensembleCMgeneral} which is equal to the conditional output state \eqref{eq:conditionalCM} also yield the optimal output logarithmic negativity.

Quite counterintuitively, however, and possibly linked to the failure of convexity, there are infinitely many choices of gains which yield the same optimal output logarithmic negativity --- including choices of gains which no longer lead to the conditional state but rather to highly mixed states. Hence, entanglement swapping illustrates the potential pitfalls of using the logarithmic negativity as a figure of merit. For example, Ref.~\cite{li05} discusses entanglement-swapping with one-sided displacements using only the output logarithmic negativity as a figure of merit. In this case, the use of the logarithmic negativity which can also be optimized by a one-sided gain (as revealed by Eq.~\eqref{eq:optGainPath}) suggests that the swapping succeeds reasonably well. Hence, the use of the logarithmic negativity effectively hides the crucial fact that one-sided gains are far from optimal if other figures of merit are considered. In this context, it is also important to note that, for symmetric two-mode Gaussian states, the logarithmic negativity is interesting from an operational point of view because it is directly linked to the optimal EPR correlations~\cite{adesso04}. However, as pointed out in Ref.~\cite{adesso04}, this connection no longer holds for asymmetric Gaussian states. Hence, even though the logarithmic negativity remains optimal if one of the asymmetric gains along the path $\vec{\gamma}(t)$ is used, no conclusion can be drawn about the EPR correlations due to the asymmetricity of the resulting states.

\section{Quantum Communication Over Lossy Channels}
\label{sec:qucomm}
The states of the form \eqref{eq:simpleCM} form a fairly general class. They include, for example, the practically relevant cases of two-mode squeezed states sent through lossy or noisy fibers. Therefore, the findings of the previous section regarding the optimal output state can be applied directly to many Gaussian quantum communication scenarios. Naturally, applications of the purity-preserving feature of the optimal Gaussian entanglement swapping come to mind. In the context of quantum communication over a lossy channel, for example, output states with higher purity (compared to direct transmission) can be obtained with the help of an intermediate entanglement swapping step: instead of sending a two-mode squeezed state with squeezing $r$ over the whole length $l$, one can send two two-mode squeezed states with squeezing $r$ over only $l/2$ such that they each incur less loss and are less mixed. Since the swapping preserves the purity, the output state in the swapping-based scenario is less mixed than the one obtained via direct transmission.

This possibility to distribute entangled Gaussian states with higher purity could be suspected to be an interesting ingredient for a continuous variable repeater scheme where higher purity might be of use, for instance, in an entanglement distillation step. Analogously to the qubit case, one might also expect benefits of entanglement swapping in continuous variable quantum key distribution if experimental imperfections are taken into account.

\begin{figure}[]
\centering
\subfigure[ The output entanglement of formation $E_F$ as a function of the input two-mode squeezing $r$ for direct transmission (black, dotted) and swapping-based transmission (blue, solid). The different pairs of curves show different total transmission lengths $l$ in units of the absorption length $l_a$; from top to bottom: $l/l_a =0.5,\,1,\,2$.]{
\includegraphics[width=0.46 \textwidth]{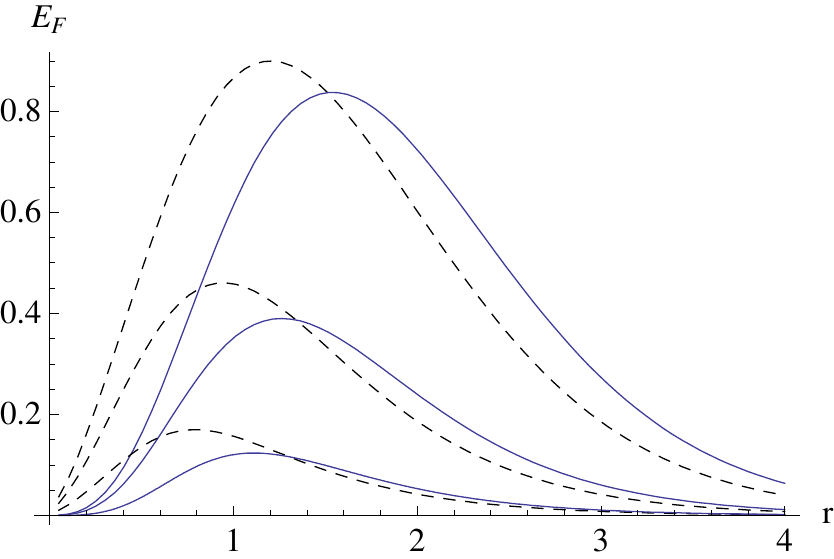}
\label{fig:eofPlot}
}
\subfigure[ The (numerically optimized) output EPR correlations $\Delta \text{EPR}_\text{opt}$ as a function of the input two-mode squeezing $r$ for direct transmission (black, dotted) and swapping-based transmission (blue, solid). The total transmission length $l$ is given by $0.5 \; l_a$ where $l_a$ is the absorption length.]{
\includegraphics[width=0.46 \textwidth]{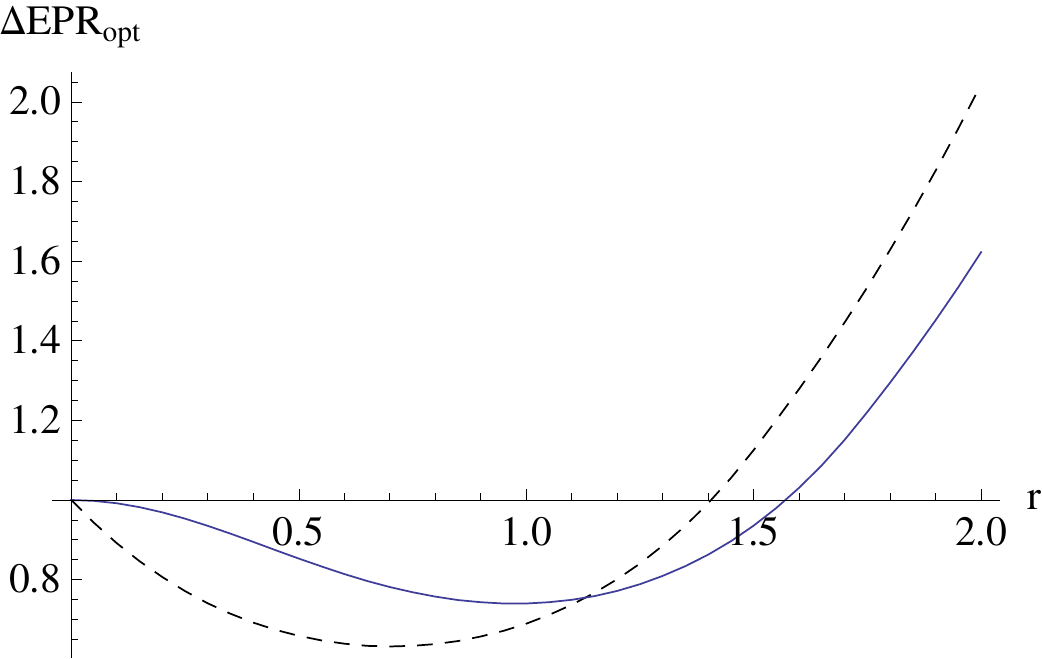}
\label{fig:eprPlot}
}
\caption{The output entanglement of formation $E_F$ and the output (optimized) EPR correlations $\Delta \text{EPR}_\text{opt}$ as a function of the input two-mode squeezing $r$. The black, dashed curves correspond to direct transmission of one mode over the whole length $l$. The blue curves correspond to transmission with an intermediate entanglement swapping step as sketched in Fig~\ref{fig:quantum_relay}. In both cases, the mode distribution is asymmetric, i.e.,~only one mode is sent through the fiber, the other mode remains at the starting point ($\tau_a=1$ in Fig~\ref{fig:quantum_relay}). For high values of $r$, the swapping-based scheme yields more entanglement of formation and better (smaller) EPR correlations. However, in general, even higher entanglement of formation and even better EPR correlations can be obtained by direct transmission if less input squeezing is used.\newline
\hspace*{\parindent}
Note that, both for the swapping-based scheme as well as for direct transmission, the output states remain entangled for any positive input squeezing. This is in fact a peculiarity of the case of asymmetric mode distribution. If both modes are subject to losses, there are values of the squeezing for which the output state is not entangled anymore.}
\label{fig:plots}
\end{figure}

In fact, numerical calculations comparing direct transmission of a two-mode squeezed state with squeezing $r$ over a lossy fiber of length $l$ to direct transmission of two copies of the same state over $l/2$ with an intermediate entanglement swapping step reveal that, for certain values of $r$, the swapping-based scheme allows to distribute more entanglement of formation and better (smaller) EPR correlations (defined in Eq.~\eqref{eq:EPRopt}) than direct transmission (see Fig.~\ref{fig:plots}). The calculation of the entanglement of formation is based on Ref.~\cite{marian98}.

\begin{figure}[]
\includegraphics[width=0.46 \textwidth]{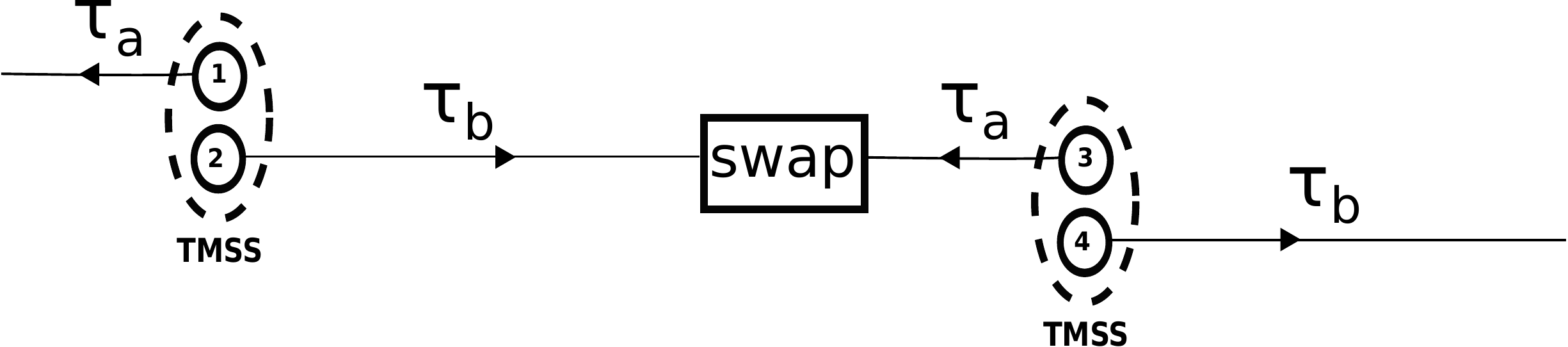}
\caption{Sketch of the swapping-based distribution scheme. Two two-mode squeezed states with squeezing $r$ are distributed over lossy channels which are characterized by transmittivities $\tau_a$ and $\tau_b$ for the first and the second mode of the two-mode squeezed states, respectively. In the middle, an entanglement swapping step is performed.}
\label{fig:quantum_relay}
\end{figure}

Note that Fig.~\ref{fig:plots} shows the output entanglement of formation only for the case of completely asymmetric mode distribution in which one mode remains fixed and the other mode is sent over the whole distance. In this case, the swapping-based scheme yields higher output entanglement of formation for high input squeezing. We find numerically that the same behavior can be observed for many different ways to split up the channel (i.e.~for many different choices of $\tau_a,\tau_b$ in Fig.~\ref{fig:quantum_relay}). A notable exception is the completely symmetric case ($\tau_a=\tau_b$ in Fig.~\ref{fig:quantum_relay}) for which we cannot find any input squeezing such that the swapping-based scheme yields higher output entanglement of formation than direct transmission.\footnote{Note also that, in general, the logarithmic negativity shows a behavior quite different from the behavior of the entanglement of formation. For direct transmission, one can show that the logarithmic negativity always (for all choices of $\tau_a,\tau_b > 0$ in Fig.~\ref{fig:quantum_relay}) grows monotonically with the input squeezing and quickly approaches a maximum value which depends on the transmission length. Numerical calculations suggest that the same is true for swapping-based transmission. We cannot find transmittivities $\tau_a,\tau_b>0$ and input squeezing $r$ such that the swapping-based scheme yields higher output logarithmic negativity than direct transmission. 
}

This result is consistent with the intuition that for large squeezing, when the swapping becomes highly efficient and the states are more sensitive to losses, entanglement swapping might help.
Does this mean that swapping can improve the performance of Gaussian entanglement distribution schemes?

To answer this question in greater generality, we consider two two-mode squeezed states with squeezing $r$ which are distributed over a general lossy channel characterized by transmittivities $\tau_a$ and $\tau_b$ for the first and second modes, respectively. The whole swapping-based distribution scheme is sketched in Fig.~\ref{fig:quantum_relay}.
The resulting states are characterized by a CM of the form \eqref{eq:simpleCM} with matrix elements
\begin{equation}
\label{eq:outputParams}
a,b= 1+ \tau_{a,b} ( \cosh(2r)-1), c = \sqrt{\tau_a \tau_b} \sinh(2r).
\end{equation}
Note that, from here on, we are using units in which the vacuum variance is given by 1, for simplicity.

Using expression \eqref{eq:optimalOutCM} we find that the output state after optimal swapping with these states is characterized by the parameters
\begin{eqnarray}
\label{eq:optCMparams}
a_\text{opt},b_\text{opt}&=& 1+ 2 \tau_{a,b} \sinh^2(r) - c_\text{opt}, \\
 c_\text{opt} &=& \frac{\tau_a \tau_b  \sinh^2(2r)}{2 - \tau_a - \tau_b +(\tau_a+\tau_b) \cosh(2r)} . \nonumber
\end{eqnarray}

\section{Effective Loss Analysis}
\label{sec:effloss}
We now show that direct transmission over a lossy channel always yields states which are at least as good as the states obtainable in the swapping-based scheme which are characterized by Eq.~\eqref{eq:optCMparams}. An effective loss analysis, which has been used before in Ref.~\cite{lund09}, reveals that the swapped states --- as long as they are still entangled --- are in fact, lossy two-mode squeezed states. More precisely, they are two-mode squeezed states with squeezing $r_\text{eff}$ whose first and second mode have been sent through lossy channels characterized by transmittivities $\tau_{a,\text{eff}}$ and $\tau_{b,\text{eff}}$, respectively. In general, the effective squeezing $r_\text{eff}$ and the effective transmittivities $\tau_{a,\text{eff}},\tau_{b,\text{eff}}$ which characterize the swapped states are different from the squeezing $r$ and the transmittivities $\tau_{a},\tau_{b}$ that characterize the input states used in the swapping. Once the effective transmittivities for the swapped states have been calculated, one can compare the total effective transmittivity $\tau_{a,\text{eff}} \, \tau_{b,\text{eff}}$ to the total transmittivity that can be achieved by direct transmission. Only if the total effective transmittivity for the swapping-based scheme is higher than that for direct transmission, will it be beneficial to use the swapping instead of direct transmission. Unfortunately, we will now show that this is never the case.

We showed that, for entanglement swapping with two-mode Gaussian states whose CM is of the form \eqref{eq:simpleCM}, the output states after optimal entanglement swapping are of the same form, but with new parameters $a,b,c$ given by Eq.~\eqref{eq:outputParams}. Note that the parameters $a,b,c$ which arise in this manner are not completely arbitrary, but are constrained by the requirement that the corresponding CM is the CM of a physical state. Before we calculate the effective transmittivities and the effective squeezing for the output states after the swapping, let us collect all such constraints on $a,b,c$.

First of all, the parameters $a,b,c$ all have to be real, and the diagonal elements $a,b$ have to be non-negative since they are the variances of the quadrature operators. Additionally, the resulting CM has to fulfill the Roberston-Schr\"odinger inequality which is equivalent to the following inequalities for $a,b,c$,
\begin{equation}
\label{eq:physicality}
(a-1) (b+1)\geq c^2, \quad (a+1) (b-1)\geq c^2.
\end{equation}
These constraints on $a,b,c$ are assumed in what follows since we are always dealing with covariance matrices of (Gaussian) quantum states.

Furthermore, we will focus on those output states of the entanglement swapping which are entangled. A two-mode Gaussian state is separable exactly if it fulfills the PPT criterion \cite{simon00}. For a physical state with a CM of the simple standard form \eqref{eq:simpleCM}, the PPT criterion is equivalent to the inequality
\begin{equation}
a+b + c^2 -1 \leq a b.
 \label{eq:separability}
\end{equation}

Now, any CM of the standard form \eqref{eq:simpleCM} can formally be written as the CM of a lossy two-mode squeezed state with parameters
\begin{eqnarray}
a,b &=& 1+ \tau_{a,b,\text{eff}} ( \cosh(2r_\text{eff})-1), \nonumber\\ c &=& \sqrt{\tau_{a,\text{eff}} \tau_{b,\text{eff}} } \sinh(2r_\text{eff}),
\end{eqnarray}
where the effective transmittivities $\tau_{a,\text{eff}}, \tau_{b,\text{eff}}$ and the effective squeezing $r_\text{eff}$ are given by
\begin{eqnarray}
&&\cosh( 2 r_\text{eff} ) = \frac{c^2  + (a-1)(b-1)}{c^2  - (a-1)(b-1)}, \\
&&\tau_{a,\text{eff}} = \frac{a -1}{\cosh( 2 r_\text{eff} ) - 1} ,\quad
\tau_{b,\text{eff}} = \frac{b -1}{\cosh( 2 r_\text{eff} ) - 1}.  \nonumber
\label{eq:effParams}
\end{eqnarray}
However, the resulting parameters $\tau_{a,\text{eff}}, \tau_{b,\text{eff}}$ and $r_\text{eff}$ can only be interpreted as transmittivities and squeezing if $0\leq \tau_{a,\text{eff}}, \, \tau_{b,\text{eff}} \leq 1$ and $1 \leq \cosh( 2 r_\text{eff} ) < \infty$. It follows from the physicality constraints in Eq.~\eqref{eq:physicality} that $a,b\geq 1$. Therefore, $1 \leq \cosh( 2 r_\text{eff} ) < \infty$ is equivalent to $c^2 > (a-1)(b-1)$ which is just the negation of the separability condition \eqref{eq:separability}. Furthermore, it is easy to show that $c^2 > (a-1)(b-1)$, together with the physicality constraints \eqref{eq:physicality}, imply $0\leq \tau_{a,\text{eff}}, \, \tau_{b,\text{eff}} \leq 1$. Hence, for physical parameters $a,b,c$, violation of the separability condition \eqref{eq:separability} is equivalent to  $1 \leq \cosh( 2 r_\text{eff} ) < \infty$ and $0\leq \tau_{a,\text{eff}},\, \tau_{b,\text{eff}} \leq 1$. Therefore, exactly those two-mode Gausian states of the form $\eqref{eq:simpleCM}$ which are entangled are equal to lossy two-mode squeezed states.

For these states, Eqs.~\eqref{eq:effParams} yield the following expressions for the effective transmittivities and the effective squeezing as a function of the squeezing $r$ and the transmittivities $\tau_a$ and $\tau_b$ of the input states used in the entanglement swapping,
\begin{widetext}
\begin{eqnarray}
\cosh( 2 r_\text{eff} ) &=& \frac{
2  \left(\tau_a  -\tau_a^2 + \tau_b - \tau_b^2 \right) \cosh (2 r) + \tau_a \tau_b \cosh (4 r)
+ 7 \tau_a \tau_b - 6 (\tau_a + \tau_b)  +  2 \left(\tau_a^2+\tau_b^2 \right) + 4
}{
2 (\tau_a + \tau_b-1)   ((\tau_a + \tau_b) \cosh (2 r)- \tau_a -  \tau_b + 2)
} , \nonumber \\
\tau_{a,\text{eff}} &=&    -\frac{2 \tau_a (\tau_a+\tau_b-1) \sinh ^2(r)}{2 \tau_a+\tau_b-\tau_b \cosh (2 r)-2}  \label{eq:effParamsOut}, \quad 
\tau_{b,\text{eff}} = -\frac{2 \tau_b (\tau_a+\tau_b-1) \sinh^2(r)}{-\cosh (2 r) \tau_a+\tau_a+2 \tau_b-2}. 
\end{eqnarray}
\end{widetext}

Let us now turn back to the swapping-based scheme as sketched in Fig.~\ref{fig:quantum_relay} and use Eqs.~\eqref{eq:effParamsOut} to compare its performance to direct transmission. Denote the total distance to be bridged by $l$. Then each of the states used in the swapping is transmitted over a lossy fiber of length $l/2$. Hence we have $\tau_a \tau_b = \exp(-l/2l_a)$ where $l_a$ denotes the absorption length for the lossy fiber.  The transmittivity for a state which is directly transmitted over the whole distance $l$, on the other hand, is given by $ \exp(-l/l_a) = \tau_a^2 \tau_b^2$. Therefore, the swapping scheme is going to lead to an improvement with regards to the transmittivity exactly if the total transmittivity for the effective state $\tau_{a,\text{eff}} \tau_{b,\text{eff}}$ is bigger than the total transmittivity $\tau_a^2 \tau_b^2$ for direct transmission. Using Eqs.~\eqref{eq:effParamsOut}, we obtain
\begin{eqnarray}
&&\tau_{a,\text{eff}} \, \tau_{b,\text{eff}}=  \\
&&\frac{4 \tau_a \tau_b (\tau_a+\tau_b-1)^2 \sinh ^4(r)}
{
((2 \cosh  r-1) \tau_a + 2 - 2\tau_b) ((2 \cosh  r-1) \tau_b+2-2 \tau_a)} . \nonumber
\label{eq:totalEffLoss}
\end{eqnarray}
One can now show that --- under the assumption that the output state is entangled, hence that it violates \eqref{eq:separability} --- the total effective transmittivity $\tau_{a,\text{eff}} \tau_{b,\text{eff}}$ for the optimal swapping \eqref{eq:totalEffLoss} is never bigger than $\tau_a^2 \tau_b^2$. For the limiting case of perfect transmission ($\tau_a= \tau_b = 1 $), also the total effective transmission $\tau_{a,\text{eff}} \tau_{b,\text{eff}}$ in the swapping-based scheme is 1. This means that, for swapping with pure two-mode squeezed states, the output state is a pure two-mode squeezed state again, as shown before in Ref.~\cite{loock02}. In general, the total effective transmittivity achieved by swapping is strictly smaller than the total transmittivity for direct transmission.

Hence direct transmission is always at least as good as swapping in the following sense: any entangled output state that can be distributed over a given length $l$ using an intermediate entanglement swapping step can be obtained \emph{identically} by direct transmission over the same or an even longer distance. Consequently, if the distance is fixed, the same or even better states are obtained by direct transmission. Since direct transmission can yield the very same states as swapping, it can outperform swapping no matter what the intended application of the distributed entanglement is. In particular, the swapping-based scheme cannot lead to an enhanced performance of continuous variable entanglement distillation. Therefore, the first step in a continuous variable repeater must be an entanglement distillation step. Furthermore, contrary to the qubit case where the use of swapping in a quantum relay scheme has been shown to be beneficial \cite{collins05}, continuous variable swapping cannot lead to improved performance in continuous variable quantum key distribution.

There remains a small puzzle with regards to the benefits of swapping for high squeezing $r$, as demonstrated in Fig.~\ref{fig:plots}. The solution to this puzzle is given by the observation that the swapped states are not only characterized by effective transmittivities which are different from the transmittivities of the input states, but also by a different effective squeezing. The fact that, even though the transmittivity does not improve, the entanglement of formation as well as the EPR correlations are sometimes better for the swapping-based scheme than for direct transmission in the high-squeezing regime is due to the fact that the swapping leads to a lower effective squeezing. In the high-squeezing regime, more squeezing does not lead to a further improvement of the output entanglement of formation and output EPR correlations since increasing squeezing means increasing sensitivity to losses and, correspondingly, increasing mixedness of the output state. This reasoning implies that direct transmission should allow the distribution of the same amount of entanglement of formation and the same EPR correlations by using input states with lower squeezing. This can in fact be observed in Fig.~\ref{fig:plots}. The plots also show that, for a fixed total transmission length, if the optimal amount of input squeezing is used, the amount of entanglement of formation that can be distributed using the optimal amount of input squeezing is even higher for direct transmission than for the swapping-based scheme. Similarly, if the optimal amount of input squeezing is used, direct transmission yields better EPR correlations than the swapping-based scheme. This is to be expected, since, as shown above, the transmittivity is lower for the swapping-based scheme.

\section{Discussion}
\label{sec:discussion}
In this paper, we discussed entanglement swapping with mixed Gaussian states. We calculated the optimal gains for a large class of mixed Gaussian states which contain the most practically relevant states, namely lossy and noisy two-mode squeezed states. These optimal gains yield in fact strictly better output states than the gains that have been used for entanglement swapping with mixed Gaussian states in \cite{liu02,ban04,li05}. In these references, only one-sided displacements have been taken into account which leads to suboptimal output states and potentially to premature negative conclusions. For example, in Ref.~\cite{liu02}, it is claimed that ``there is more [EPR] noise resulting from the entanglement swapping than [...] from direct transmission, regardless of the values of the displacement gain.'' This is only true if only one-sided displacements are taken into account. In fact, as demonstrated by Fig.~\ref{fig:eprPlot}, there are values of the input squeezing $r$ for which swapping does yield better EPR correlations than direct transmission.

Therefore a direct comparison of the output states of swapping-based schemes with the output states of direct transmission for fixed input squeezing $r$ suggests that swapping can in fact be beneficial; in any case, it does not yet warrant the negative conclusions that have previously been drawn. To assess the benefits of swapping in quantum communication in greater generality, we performed an effective loss analysis of the swapping-based scheme in the second part of the paper. We showed that the output states of the optimal entanglement swapping --- as long as they are entangled --- are always equal to lossy two-mode squeezed states characterized by an effective squeezing $r_\text{eff}$ and effective transmittivities $\tau_{a,\text{eff}}$ and $\tau_{b,\text{eff}}$. This implies also that direct transmission over a lossy fiber can be mimicked by entanglement swapping. Then, we calculated the effective squeezing $r_\text{eff}$ and effective transmittivities $\tau_{a,\text{eff}}$ and $\tau_{b,\text{eff}}$ for the swapped states as a function of the input squeezing and the input transmittivities and showed that the total effective transmittivity for the swapped state is never bigger than the total transmittivity for direct transmission. This implies that direct transmission yields the same or better output states than swapping-based schemes. In cases where swapping seems to achieve better results than direct transmission, such as in Fig.~\ref{fig:plots}, the advantage of swapping results from an effective squeezing which is different from the input squeezing. Therefore, the same or better results can be achieved by using direct transmission and a different value of the input squeezing.

A caveat remains since we only analyzed the case in which the swapping is performed in the middle of the section to be bridged. There still remains the possibility that an asymmetric position of the swapping station, say after 1/3 of the total distance yields better results. Intuitively, there is good reason not to expect it. The extreme case of such an asymmetric scenario would be given by a scenario in which state 1 remains at the same place and state 2 travels over the whole distance $l$. Hence, state 2 would evolve into a lossy two-mode squeezed state $\rho_l$ which is then swapped via the pure, finitely squeezed two-mode squeezed state 1. In this case, the known results for teleportation via pure finitely squeezed states imply that the Wigner function of the output state is the convolution of the Wigner function of the state $\rho_l$ with a Gaussian distribution whose width is inverse to the squeezing of state 1~\cite{braunstein98}. Hence, the only effect of the final swapping step is a further broadening of the variances of the state $\rho_l$. In the case of direct transmission, on the other hand, we would also obtain the state $\rho_l$, and it would obviously be better to use that state directly.
\vspace*{1em}

\begin{acknowledgments}
J.H. acknowledges support through the BayBFG and the Studienstiftung des deutschen Volkes.
P.v.L. acknowledges the DFG for financial support through the Emmy Noether programme.
\end{acknowledgments}


\begin{thebibliography}{99}
\bibitem{zukowski93}
M. \.{Z}ukowski, A. Zeilinger, M. A. Horne, and A. K. Ekert, Phys. Rev. Lett. \textbf{71}, 4287--4290 (1993).

\bibitem{briegel98}
H.-J.  Briegel, and W. D\"ur, J. I. Cirac, and P. Zoller, Phys. Rev. Lett. \textbf{81}, 5932 (1998).

\bibitem{sen05}
A. Sen(De), U. Sen, C. Brukner, V. Bu\v{z}ek, and M. \.{Z}ukowski, Phys. Rev. A \textbf{72}, 042310 (2005).

\bibitem{modlawska08}
J. Modlawska and A. Grudka, Phys. Rev. A \textbf{78}, 032321 (2008).

\bibitem{collins05}
D. Collins, N. Gisin, and H. De Riedmatten, Journal of Modern Optics \textbf{52(5)}, 735-753 (2005).

\bibitem{eisert03}
J. Eisert and M. Plenio, Int. J. Quant. Inf. \textbf{1}, 479 (2003).

\bibitem{adesso07}
G. Adesso and F. Illuminati, J. Phys. A: Math. Theor. \textbf{40}7821 (2007).

\bibitem{braunstein05}
S. Braunstein and P. van Loock, Rev. Mod. Phys. \textbf{77}, 513--577 (2005).

\bibitem{pirandola06}
S. Pirandola, D. Vitali, P. Tombesi, and S. Lloyd, Phys. Rev. Lett. \textbf{97}, 150403 (2006).

\bibitem{liu02}
{J.-M.} {Liu}, J. {Li}, and {G.-C.} {Guo}, Chinese Physics \textbf{11}, 339--345 (2002).

\bibitem{ban04}
M.~Ban, Journal of Physics A \textbf{37}, 31 L385--L390 (2004).

\bibitem{li05}
H. Li, F. Li, Y. Yang, and Q. Zhang, Phys. Rev. A 71, 022314 (2005)

\bibitem{loock02}
P. {van Loock}, Fortschritte der Physik \textbf{50}, 12 1177--1372 (2002).

\bibitem{einstein35}
A. Einstein, B. Podolski, and N. Rosen, Phys. Rev. \textbf{47}, 777 (1935).

\bibitem{braunstein98}
S. L. Braunstein and H. J. Kimble, Phys. Rev. Lett. \textbf{80}, 869--872 (1998).

\bibitem{furusawa98}
A. Furusawa, J. L. Sørensen, S. L. Braunstein, C. A. Fuchs, H. J. Kimble, and E. S. Polzik, Science \textbf{282} no. 5389, pp. 706--709 (1998).

\bibitem{simon00}
R. Simon, Phys. Rev. Lett. \textbf{84} 2726 (2000).

\bibitem{duan00}
L. M. Duan, G. Giedke, J. I. Cirac, and P. Zoller, Phys. Rev. Lett. \textbf{84} 2722 (2000).

\bibitem{eisert01}
J. Eisert, Ph.D. thesis, University of Potsdam (2001).

\bibitem{vidal02}
G. Vidal, and R. F. Werner, Phys. Rev. A \textbf{65}, 32314 (2002).

\bibitem{plenio05}
M. B. Plenio, Phys. Rev. Lett. \textbf{95}, 090503 (2005).

\bibitem{adesso04}
G. Adesso, A. Serafini, and F. Illuminati, Phys. Rev. A \textbf{70}, 022318 (2004).


\bibitem{marian98}
P. Marian and T. A. Marian, Phys. Rev. Lett. \textbf{101}, 220403 (2008).

\bibitem{lund09}
A. P. Lund, and T.C. Ralph, Phys. Rev. A \textbf{80}, 032309 (2009)

\end{thebibliography}
\end{document}